\documentclass[12pt]{article}
\usepackage{epsfig,multicol,latexsym,amsfonts,amssymb}
\usepackage{amsmath}
\PassOptionsToPackage{righttag}{amsmath}
\def\tntnbar{${\bf10}\oplus{\bf\overline{10}}$}

\def\fpi{$\{\pi\}$}
\def\frho{$\{\rho\}$}
\def\fbOne{$\{b_1\}$}
\def\faOne{$\{a_1\}$}
\def\faTwo{$\{a_2\}$}
\newdimen\digitwidth
\setbox0=\hbox{\rm0}
\digitwidth=\wd0
\def\ftm{\phantom}
\def\indnt{\vskip0pt\noindent\vsp\hskip1.5em}
\def\vsp{\hbox{\vrule height12.5pt depth3.5pt width0pt}}
\def\sqr#1#2{{\vcenter{\vbox{\hrule height .#2pt
        \hbox{\vrule width .#2pt height#1pt\kern#1pt
                   \vrule width.#2pt}\hrule height.#2pt}}}}

\def\brlist{\begin{thebibliography}{999}}
\def\brf{\bibitem}
\def\erlist{\end{thebibliography}}
\def\bra{\langle}
\def\ket{\rangle}
\def\eqn#1{(\ref{#1})}
\def\jtem#1#2{\par\hangafter0\hangindent#1
              \noindent\llap{#2\enspace}\ignorespaces}
\def\tbset#1{${}^{\,\rm #1\,}$}

\def\tbnote#1#2#3{\vskip2pt\noindent\jtem{12pt}{$\ {}^{\rm #1\,}$}
    \hsize=#2{\def\baselinestretch{1.0}\small #3\vskip0pt}}

\catcode`@=11
\def\eqalign#1{\null\,\vcenter{\openup\jot\m@th
  \ialign{\strut\hfil$\displaystyle{##}$&$\displaystyle{{}##}$\hfil
      \crcr#1\crcr}}\,}
\def\eqalignno#1{\displ@y \tabskip\@centering
  \halign to\displaywidth{\hfil$\@lign\displaystyle{##}$\tabskip\z@skip
    &$\@lign\displaystyle{{}##}$\hfil\tabskip\@centering
    &\llap{$\@lign##$}\tabskip\z@skip\crcr
    #1\crcr}}
\def\leqalignno#1{\displ@y \tabskip\@centering
  \halign to\displaywidth{\hfil$\@lign\displaystyle{##}$\tabskip\z@skip
    &$\@lign\displaystyle{{}##}$\hfil\tabskip\@centering
    &\kern-\displaywidth\rlap{$\@lign##$}\tabskip\displaywidth\crcr
    #1\crcr}}

\long\def\@makefntext#1{
  \vskip0pt\parindent0pt\begin{list}{}%
  {\labelwidth1.5em\leftmargin=\labelwidth%
     \labelsep3pt\itemsep0pt\parsep0pt\topsep-2pt
            \def\baselinestretch{1.0}\footnotesize}%
  \item[\hfill\@makefnmark]#1\end{list}}

\long\def\@makecaption#1#2{\vskip10pt
    {#1:\ \begingroup\small\baselineskip14pt plus4pt minus2pt
             #2\par\endgroup}}

\catcode`@=12
\def\bln#1#2\eln{\begin{equation}\label{#1}%
   \eqalign{#2}\end{equation}\vskip3mm\noindent}
\def\bqt#1#2\eqt{\addtocounter{equation}{0}
	$$\eqalignno{\refstepcounter{equation}\label{#1}#2}$$}
\def\eql#1#2{\hbox{(\ref{#1}$#2$)}}
\def\mbb#1{\mathbb #1}	
\def\mbf#1{\mbox{\boldmath $#1$}}

\def\sct{\scriptstyle}

\def\ocal{\hskip0.6em\raise0.70em\hbox{$\sct\circ$}\kern-0.94em}
\def\ocap{\hskip0.2em\raise0.80em\hbox{$\sct\circ$}\kern-0.50em}
\begin{document}
\begin{figure}[ht]\vglue-1mm
\begin{center}\mbox{\epsfig{file=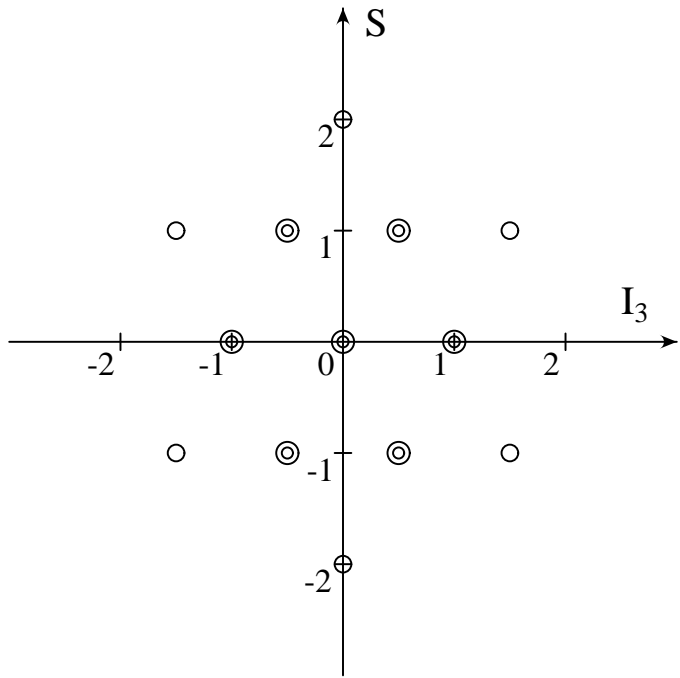,width=3.5in}}
		\end{center}\vskip1mm
\caption{Weight Diagram for multiplet \tntnbar; 
	single circles 
	have just one member
of the multiplet, while the double circles indicate two occupancies
by the members of the multiplet.}
\label{fg01}\end{figure}

\thispagestyle{empty}
\begin{center}{\sf BONN/HISKP REPORT}\end{center}
\vglue 2mm
%
\rightline{BONN-HISKP-03-0202}
\vskip 2cm
\begin{center}{\Large\bf
Vector mesons in $q\bar qq\bar q$ systems\\}
\vskip 24pt
S. U. Chung$^{\ *}$
\vskip 2pt
{\em Physics Department, 
Brookhaven National Laboratory, Upton, NY 11973}
\vskip 12mm
E. Klempt
\vskip 2mm
{\em Helmholtz-Institut f\"ur Strahlen- und Kernphysik\\
 Universit\"at Bonn, Nu{\ss}allee 14--16, D-53115 Bonn, Germany}
\vskip15mm
\today
\vskip 3cm
{\large\bf abstract}
\vskip 1cm
\begin{quote}
\indnt
   We discuss the vector mesons, 
$I^G(J^{PC})=1^+(1^{--})$ and $1^-(1^{-+})$ in \tntnbar\ representations,
decaying into two ground-state octets.
We derive a powerful selection rule, valid in the limit of
flavor $SU(3)$ symmetry.  The octets considered are 
\fpi, \frho, \fbOne, \faOne\ and \faTwo, labeled by the isovector state
in the representation.
\vtop{\hbox{Nonstrange vector mesons; Four-quark exotic mesons;}
	\hbox{$SU(3)$ symmetry.}}
\end{quote}
\vskip3mm
\fbox{\large\bf To be Published in Phys. Lett. B}
\end{center}
\vspace*{\fill}\footnoterule\vskip3pt\footins{\small$^*$
Mercator Visiting Professor at Technische Universit\"at M\"unchen 
and Universit\"at Bonn.}
\eject
\pagenumbering{arabic}
\def\baselinestretch{1.4}\normalsize
\pagenumbering{arabic}
\def\baselinestretch{1.4}\normalsize
%
%
\indnt
   An exotic meson, the $\pi_1(1400)$ with $I^G(J^{PC})=1^-(1^{-+})$,
has been seen to decay into a $p$-wave $\eta\pi$ system~\cite{pi1a}. 
It has been shown~\cite{su3a} that, if the $\eta$ meson is assumed 
to be a (pure) member of the pion octet and in the limit of 
flavor $SU(3)$ conservation in its decay, a $p$-wave 
$\eta\pi$ system belongs to a flavor \tntnbar\ 
representation by the requirement of Bose symmetrization.
This implies that the $\pi_1(1400)$ must belong to a family of four-quark
states ($q\bar q+q\bar q$). In contrast, a second exotic meson,
the $\pi_1(1600)$ with $I^G(J^{PC})=1^-(1^{-+})$, is reported to have
a substantial decay mode into $\eta'\pi$~\cite{pi1b}.  
Whatever its constituent quark content may be---since the 
$\eta'$ is mostly an $SU(3)$ singlet---it is a member 
of an octet.
\indnt
   The multiplet \tntnbar\  
is given in a pictorial form in Fig.\ref{fg01}.\vskip0pt
\vskip0pt\noindent
The purpose of this letter is explore the consequence of the existence
of other vector mesons predicted to exist 
in the multiplet \tntnbar~\cite{su3a}.
In the limit of flavor $SU(3)$ symmetry, {\em the mass of the 
$I^G(J^{PC})=1^+(1^{--})$ member 
must be equal to that of the $\pi_1(1400)$.} 
We shall adopt the notation $\rho_x(1400)$ for this state.
We know that {\em both} must belong to the $q\bar qq\bar q$ family of mesons.

   A full account of the $J^{PC}=1^{--}$ and $1^{-+}$
`vector' mesons with $q\bar qq\bar q$ entails existence of
{\em two} sets of `supermultiplets' with 81 members\footnote{The number
of states in a supermultiplet can derived in two different ways:
$\{q\bar q+q\bar q\}=({\bf1}\oplus{\bf8})\otimes({\bf1}\oplus{\bf8})$ or
$\{qq+\bar q\bar q\}=(\overline{\bf3}\oplus{\bf6})\otimes
			({\bf3}\oplus\overline{\bf6})$.}
in each.  Each supermultiplet has the structure
\bln{i01}
\mbb{V}=\{q\bar qq\bar q\}
	&=2\times{\bf1}\ \oplus\ 4\times{\bf8}\ \oplus\ 
	{\bf10}\ \oplus\ \overline{\bf10}\ \oplus\ {\bf27}
\eln
where we consider here three flavors $q=\{u,d,s\}$.
Of course, not all members of the multiplets need to support quasi-bound-state
systems---with reasonably finite widths---which can be 
identified experimentally.

   Let $\chi^0$ be the wave function for a nonstrange, 
neutral state in any of the representations listed above 
(there is at least one such state in each).
We now define the supermultiplets through the
charge-conjugation operator $\mbb{C}$ 
\bln{i01a}
\mbb{V}_\zeta:\hskip6mm
   &\mbb{C}\,|\chi^0(\mbf{n})\ket=\zeta\,|\chi^0(\mbf{n})\ket,\qquad
	\mbf{n}={\bf1},\ {\bf8}, \ {\bf27}\cr
&\mbb{C}\,|\chi^0({\bf10})\ket=\zeta\,|\chi^0(\overline{\bf10})\ket,\quad
 \mbb{C}\,|\chi^0(\overline{\bf10})\ket=\zeta\,|\chi^0({\bf10})\ket
\eln
where $\zeta=\pm1$.  Thus, $\zeta$ is simply the $\mbb{C}$ eigenvalue
of a $\chi^0$ which belongs to any of the `self-conjugate' 
representations ${\bf1}$, ${\bf8}$ and ${\bf27}$. 
The subscript therefore refers
to the $C$-parity of the dominant $J^{PC}$ in the family ($\zeta=C$).
Each $\mbb{V}_\pm$ supermultiplet comes with 61 $J^{PC}=1^{-\pm}$
members belonging to the self-conjugate representations of $SU(3)$. 
The \tntnbar\  members are broken up into 14 $J^P=1^-$ strange members 
and {\em two} sets of three $I^G(J^{PC})=1^\pm(1^{-\mp})$ members.
Thus the physical states in eigenstates of $\mbb{C}$ are
\bln{cf6a}
   \chi^0_{\pm}={1\over\sqrt{2}}\,
\left[\chi^0({\bf10})\pm\zeta\,\chi^0(\overline{\bf10})\right],\quad
	\mbb{C}|\chi^0_{\pm}\ket=\pm|\chi^0_{\pm}\ket
\eln
The `quantum number' $\zeta$ has {\em nothing}
to do with the $C$-parity of the nonstrange members of \tntnbar.
We assume that one supermultiplet, $\mbb{V}_+$ or $\mbb{V}_-$, gives rise to
$\pi_1(1400)$ and $\rho_x(1400)$; we shall designate their counterparts
by $\pi'_1(1400?)$ and $\rho'_x(1400?)$ in the other supermultiplet.  
All four states are isospin triplets and
belong to \tntnbar\ representations.

   Let $X$ stand for a nonstrange state belonging 
to  a \tntnbar\ representation.
The main purpose of this paper is to point out that
the $\zeta$'s, as defined through \eqn{i01a}, 
lead to a powerful selection rule regarding
the decay $X\to a+b$ where $a$ and $b$ are members of the octets (not
necessarily the same). For the purpose, consider the decay modes 
of a $J^P=1^-$ nonstrange meson $X$ as outlined below:
\bqt{cf3}
  X&\to \left[\hbox{\fpi\fpi}\right]_P,\
	\left[\hbox{\frho\frho}\right]_{P,\,F},\
	\left[\hbox{\faOne\fpi}\right]_{S,\,D},\
	\left[\hbox{\faTwo\fpi}\right]_{D},&\eql{cf3}{a}\cr
    &\to\left[\hbox{\frho\fpi}\right]_P,
	\left[\hbox{\fbOne\fpi}\right]_{S,\,D}&\eql{cf3}{b}\cr
\eqt
where the bracket $\{\ftm{a}\}$ refers to the totality of an octet.
For example, \fpi\ stands for $\pi$, $K$, $\bar K$ and $[\eta]_8$.
(we use $[\ ]_8$ to denote the octet component of a particle; in addition,
we follow a common practice of writing $[a\,b]_L$ to indicate
an orbital angular momentum $L$ between $a$ and $b$.)
The $SU(3)$ wave functions $\phi$ (real by definition)
for nonstrange neutral members
of $\{a\}\,\{b\}$\cite{dSw} are transformed,\footnote{
There is an overall phase $\eta$ 
in the defining formula (8.2) of Ref.\cite{dSw}, which has been set to $+1$.
However, this is really not necessary for our purposes, as it can be
absorbed into $\kappa$ of \eqn{cf7}.} under $\mbb{C}$,
\bln{cf5a}
  \mbb{C}\,\phi(\overline{\bf10})=g\,\phi({\bf10}),\quad
  \mbb{C}\,\phi({\bf10})=g\,\phi(\overline{\bf10})
\eln
where $g=C_a\cdot C_b=\pm1$. 
$C$ refers to the $C$-parity of the octets involved in the final state.
Hence $g$ is equal to a product of two $G$-parities 
for the {\em isovector} members of $\{a\}$ and $\{b\}$.  
For example, $g=+1$ for the decay \fpi\fpi\ and likewise 
for all the final states listed in  \eql{cf3}{a},
while $g=-1$ for \eql{cf3}{b}.

   Because of $C$-parity conservation, the amplitudes
for the decay $X\to a+b$ must obey the relationship
\bln{cf7}
\bra\phi(\overline{\bf10})|{\cal M}|\chi^0(\overline{\bf10})\ket
	=\kappa\,\bra\phi({\bf10})|{\cal M}|\chi^0({\bf10})\ket
\eln
where $\kappa=\pm1$. (The $C$-parity conservation requires that 
the squares of the amplitudes be the same.)  We now
insert $\mbb{C}^{\,\dagger}\,\mbb{C}=I$
next to the operator ${\cal M}$ on the left side of \eqn{cf7} 
and use \eqn{i01a} and \eqn{cf5a}---to obtain $\zeta=\kappa\,g$.
{\em This is our main result}.  It can be stated as follows:
the phase $\zeta$, which determines to which supermultiplet $X$ belongs,
specifies how it should couple to a decay
product consisting of two octets.  
Our selection rule can be succinctly stated
\bln{cf8}
   X&\to\left[\{a\}\,\{b\}\right]_+, 	\qquad  
	X\not\to\left[\{a\}\,\{b\}\right]_-\cr
   X'&\to\left[\{a\}\,\{b\}\right]_-, 	\qquad  
	X'\not\to\left[\{a\}\,\{b\}\right]_+\cr
\eln
where $X\in\mbb{V}_+$ and $X'\in\mbb{V}_-$ if $\kappa=+1$ and
{\em vice versa} if $\kappa=-1$.
The subscript to $\{a\}\,\{b\}$ stand for $g=C_a\cdot C_b$.  
In another words,
the allowed decay modes for $X\to a+b$ must obey the rule
$\zeta\cdot C_a\cdot C_b=+1$ in the limit of $SU(3)$ symmetry, if $\kappa=+1$.
But a state $X$ {\em cannot} come with both $\kappa=+1$ and $\kappa=-1$ 
at the same time.  So the selection rule \eqn{cf8}
should be true always, except that there is no {\em a priori} way of knowing
to which supermultiplet the $X$ belongs.

   Let $\chi$ be the wave function for the isovector meson
$X$ in {\em any} charge state. Because $X$ and  $X'$ are isovectors,
the $\chi$'s transform under the $G$-parity operator
\bln{gdf}
&\mbb{G}\,|\chi({\bf10})\ket=-\zeta\,|\chi(\overline{\bf10})\ket,\quad
 \mbb{G}\,|\chi(\overline{\bf10})\ket=-\zeta\,|\chi({\bf10})\ket,\cr
   &\chi_{\pm}={1\over\sqrt{2}}\,
\left[\chi({\bf10})\mp\zeta\,\chi(\overline{\bf10})\right],\quad
	\mbb{G}|\chi_{\pm}\ket=\pm|\chi_{\pm}\ket
\eln
Four $G$-parity eigenstates, together with their
allowed decay modes, 
are summarized in Tables Ia and Ib (for $\kappa=+1$). 
\vskip8mm
\def\arraystretch{1.3}
\begin{minipage}[h]{14cm}
\begin{center}
\begin{tabular}{|c|c||c|c|c|c|}
\hline
\multicolumn{6}{|c|}{Table Ia:\hspace{12pt}
Decay Modes\tbset{a} for $X\,(\zeta=+1)\to\left[\{a\}\,\{b\}\right]_+$}\\
\hline\hline
$I^G(J^{PC})$ &$\sqrt{2}\,\chi_\pm$ & \fpi\fpi\tbset{b} & \frho\frho\tbset{b} &
	\faOne\fpi\tbset{c} & \faTwo\fpi \\
\hline
$1^+(1^{--})$ & $\chi({\bf10})-\chi(\overline{\bf10})$ &
$\pi\,\pi$, $K\bar K$ & $\rho\rho$, $K^*\bar K^*$ & 
$a_1\,\pi$, $K_{_{1A}}\bar K$ & $a_2\,\pi$, $K^*_{2}\bar K$ \\
$1^-(1^{-+})$ & $\chi({\bf10})+\chi(\overline{\bf10})$ &
$\pi[\eta]_8$ & $\rho[\omega]_8$ &  $[f_1]_8\,\pi$ & $[f_2]_8\,\pi$ \\
\hline
\end{tabular}
\end{center}
\vskip2mm
\begin{center}
\begin{tabular}{|c|c||c|c|}
\hline
\multicolumn{4}{|c|}{Table Ib:\hspace{12pt}
Decay Modes\tbset{a} for $X'\,(\zeta=-1)\to\left[\{a\}\,\{b\}\right]_-$}\\
\hline\hline
$I^G(J^{PC})$ &$\sqrt{2}\,\chi^{\,\prime}_\pm$ & 
	\frho\fpi\tbset{b} & \fbOne\fpi\tbset{b}\tbset{c}\\
\hline
$1^+(1^{--})$ & $\chi^{\,\prime}({\bf10})+\chi^{\,\prime}(\overline{\bf10})$ &
$[\omega]_8\,\pi$ & $[h_1]_8\,\pi$\\
$1^-(1^{-+})$ & $\chi^{\,\prime}({\bf10})-\chi^{\,\prime}(\overline{\bf10})$ &
$\rho\,\pi$, $K^*\bar K$ & $b_1\,\pi$, $K_{_{1B}}\bar K$\\
\hline
\end{tabular}
\end{center}
\tbnote{a}{14cm}{When both $n\bar n$ ($n=\{u,d\}$)
	and $s\bar s$ decays are 	allowed, \\
	the predicted branching ratio is 
	$B(n\bar n)/B(s\bar s)=1/2$.}
\tbnote{b}{14cm}{The subscripts 8 denote the `octet' component.}
\tbnote{c}{14cm}{The $K_{1A}$ and $K_{1B}$ are nearly equal
	mixtures of the $K_1(1270)$ and $K_1(1400)$.}
\end{minipage}
\vskip18pt\noindent
It should be understood that, for each $J^{PC}$, an allowed decay of $X$
(table Ia) is forbidden for $X'$ (table Ib) and {\em vice versa}.
\indnt
   A further insight can be gained by considering the transformation
properties of the wave functions in `self-conjugate' representations.
Such a wave function $\chi^0(\mbf{n})$
for nonstrange, neutral members has already
been given in \eqn{i01a}, where $\mbf{n}={\bf1}$, ${\bf8}$ or ${\bf27}$.
The $\chi^0(\mbf{n})$'s are $\mbb{C}$ eigenstates with the eigenvalues
$\zeta$.  Now let it decay into $\{a\}\,\{b\}$.
Then, under $\mbb{C}$, the final-state wave function 
$\phi(\mbf{n})$ must transform according to \eqn{cf5a},
where ${\bf10}$ and $\overline{\bf10}$ are now replaced by $\mbf{n}$
\bln{sc1}
  \mbb{C}\,|\phi(\mbf{n})\ket=g\,|\phi(\mbf{n})\ket
\eln
The eigenvalue is again given by $g=C_a\cdot C_b=\pm1$.
Consider the decay matrix element
\bln{sc2}
\bra\phi(\mbf{n})|{\cal M}|\chi^0(\mbf{n})\ket
\eln
and carry out the same process applied to \eqn{cf7}.
We then conclude that $\zeta\cdot C_a\cdot C_b=+1$.
\footnote{Note that
replacing ${\bf10}$ and $\overline{\bf10}$ by $\mbf{n}$ in \eqn{cf7}
leads to a trivial identity with $\kappa=+1$.  However, the result
$\zeta=g$ for self-conjugate representations relies on the phase convention 
$\eta=+1$ given in (8.2) of Ref.~\cite{dSw}.}
So an $X$ in a self-conjugate representation $\mbf{n}$ obeys the same
selection rule it does in \tntnbar\ representations with $\kappa=+1$.
The phase $\zeta$ is the $C$-parity of a representation $\mbf{n}$,
whereas the $\zeta$ in \tntnbar\ representations imposes conditions on
the allowed decays of $X$ but {\em not} its $C$-parity.
\indnt
   In summary, we expect two states of 
$I^G(J^{PC})=1^+(1^{-+})$ mesons, $\pi_1(1400)$ 
and $\pi'_1(1400?)$, and their vector partners 
with $I^G(J^{PC})=1^-(1^{--})$, $\rho_x(1400)$ 
and $\rho'_x(1400?)$, all belonging to \tntnbar\ 
representations of flavor $SU(3)$.  The $\pi_1(1400)$ and $\rho_x(1400)$
belong in the supermultiplet $\mbb{V}_\pm$, whereas the $\rho'_x(1400?)$
and $\pi'_1(1400?)$ are members of $\mbb{V}_\mp$.
Since the $\pi_1(1400)$ is seen to decay into $\eta\pi$, 
its partner $\rho_x(1400)$ should decay into
$\pi\pi$ and $K\bar K$, with a branching ratio $B(\pi\pi)/B(K\bar K)=1/2$.
In addition, the decay $\pi_1(1400)\to\eta\pi$ implies that the same state
{\em cannot} decay into $\rho\pi$ in the limit of $SU(3)$.
The Crystal Barrel Collaboration reported\cite{CB1}
observation of a $I^G(J^{PC})=1^+(1^{-+})$ 
state\footnote{The results are consistent with a single state with
mass at $\sim1450$ MeV and its width fixed at 310 MeV; however, there are
in fact indications of two states at 1405 and 1650 MeV.}
decaying into $\rho\pi$.
This could be the $\pi'_1(1400?)$ in the \tntnbar\ 
representation in $\mbb{V}_\mp$ or a member of 
one of the $J^{PC}=1^{-+}$ octets in $\mbb{V}_\pm$.

   Finally, as an illustration of the selection rule,
consider production of the $\pi_1(1400)$ in two charge modes
\bqt{ax0}
   \pi^-p&\to\pi^-_1(1400)\,p\to\eta\pi^-\,p&\eql{ax0}{a}\cr
   \pi^-p&\to\pi^0_1(1400)\,n\to\eta\pi^0\,n&\eql{ax0}{b}\cr
\eqt
It has been shown\cite{pi1a} that Reaction \eql{ax0}{a} is mediated by
natural-parity exchange, meaning that the exchanged Reggeons (neutral) 
should correspond to $\rho^0$, $f_2$ or the Pomeron.  We see that
Reaction \eql{ax0}{a} can proceed via $f_2$ only, since
the $\pi_1(1400)$ is not supposed to couple to $\rho\pi$ in the
limit of flavor $SU(3)$ (see Table Ia) and the Pomeron is an $SU(3)$
flavor singlet.  Reaction \eql{ax0}{b}~\cite{sdv}, on the other hand, 
requires exchange of charged Reggeons.
But the charged $\rho$, for production by natural-parity exchange, 
must be suppressed for the same
reason stated above.  Moreover, Table Ia shows that 
the $\pi_1(1400)$ production is not possible 
via unnatural-parity exchange for Reaction \eql{ax0}{b}.
The $\pi'_1(1400?)$ production is possible via natural- and unnatural-parity
exchange (i.e. $\rho$ and $b_1$ exchange---see Table Ib); but then
it should {\em not} be decaying to $\eta\pi$ in the limit of $SU(3)$.
We see that the selection rule identified in this paper has 
important experimental consequences.  However, a note of caution should
be given here; we do not know the mass difference 
of the vector mesons in $\mbb{V}\pm$. 
But, if they are sufficiently close, they will mix, and this may
weaken the selection rules.

   We have thus identified, for the first time, a powerful selection rule
for an exotic meson belonging to \tntnbar\ representations.
In a follow-up article, we will further discuss experimental ramifications
of the four exotic states in these representations.
\section*{Acknowledgment}
\indnt
   We gratefully acknowledge helpful conversations with T. Barnes/Oak Ridge,
B. Metsch/Bonn and H. Petry/Bonn.
We are indebted to L. Trueman/BNL and R. Hackenburg/BNL for
their very useful comments.
\vfil\eject
\brlist
\brf{pi1a}  D. R. Thompson {\it et al.}, 
	Phys. Rev. Lett. {\bf 79} (1997) 1630;
	S. U. Chung {\it et al.}, Phys. Rev. D {\bf 60} (1999) 092001;
	A. Abele {\it et al.}, Phys. Lett. {\bf B423} (1998)  175;
            A. Abele {\it et al.}, Phys. Lett. {\bf B446} (1999) 349.
\brf{su3a} S. U. Chung, E. Klempt and J. G. K\"orner,
		Eur. Phys. J. A {\bf 15} (2002) 539.\\
	The reader may consult the references therein for a comprehensive
	list of previous publications on this topic.
\brf{pi1b} G. M. Beladidze {\it et al.}, Phys. Lett. {\bf B313} (1993) 276;
	A. Zaitsev, {\em Proc. Eighth International Conf. on Hadron
       Spectroscopy, Beijing, China (1999),}\ edited 
            by W. G. Li, Y. Z. Huang and B. S. Zou, 
		Nucl. Phys. A675 (2000) 155c;
	E. Ivanov {\it et al.}, Phys. Rev. Lett. {\bf 86} (2001) 3977.
\brf{dSw} J. J. de Swart, Rev. Mod. Phys. {\bf 35} (1963) 916.
\brf{CB1} F. Meyer-Wildhagen, Proc. PANIC2002, 
	to be published in Nucl. Phys. A.
\brf{sdv} S. A. Sadovsky, {\em Proc. Fifth Biennial Conf. on Low-Energy
	Anitproton Physics, Villasimius, Italy (1998),}\ edited by
	C. Cical\`o, A. De Falco, G. Puddu and S. Serci,
	Nucl. Phys. A655 (1999) 131c.
\erlist
\end{document}